\documentclass[letter,scriptaddress,prl   
]{revtex4}
\usepackage{amsmath}
\usepackage{graphicx}
\usepackage{dcolumn}
\usepackage{bm}
\usepackage{graphicx}
\usepackage{epsfig}
\usepackage[usenames]{color}
\usepackage{tikz}
\usetikzlibrary{shapes}

\usepackage{setspace}
\usepackage[mathcal]{eucal}

\usepackage{subfigure}  
\usepackage{amsthm}
\usepackage{float}

\newcommand{\df}[2]{\displaystyle\frac{#1}{#2}}

\newcommand{\be}{\begin{eqnarray}}
\newcommand{\en}{\end{eqnarray}}

\newcommand{\bs}{\bigskip}
\newcommand{\ms}{\medskip}
\newcommand{\tcr}{\textcolor{red}}
\newcommand{\tcb}{\textcolor{blue}}

\newcommand{\tcm}{\textcolor{magenta}}

\begin{document}
\title{{\Large Radius Evolution for Bubbles with Elastic Shells}} 
   \author{Stefan C. Mancas}
{\email{mancass@erau.edu}
\affiliation{Embry-Riddle Aeronautical University,\\ Daytona Beach, FL. 32114-3900, U.S.A.}
\author{Haret C. Rosu}
\email{hcr@ipicyt.edu.mx}
\affiliation{Instituto Potosino de Investigacion Cientifica y Tecnologica,\\
Camino a la presa San Jos\'e 2055, Col. Lomas 4a Secci\'on, 78216 San Luis Potos\'{\i}, S.L.P., Mexico}
\author{Chun-Chung Hsieh}
\email{cchsieh@gate.sinica.edu.tw}
\affiliation{Institute of Mathematics, Academia Sinica, Nankang, Taipei 115, Taiwan}


\begin{abstract}

\noindent We present an analysis of an extended Rayleigh-Plesset (RP) equation for a three dimensional cell of
microorganisms such as  bacteria or  viruses in some liquid, where the  cell membrane in bacteria or the envelope (capsid)
in viruses possess elastic properties. To account for rapid changes in the shape  configuration of such  microorganisms,
the bubble membrane/envelope  must be rigid to resist large pressures while being flexible to adapt to growth or decay.
Such properties are embedded in the RP equation by including a pressure bending term that is proportional
to the square of the curvature of the elastic wall. Analytical solutions to this extended equation are obtained in terms
of elliptic functions.

\ms

\noindent \tcr{{\scriptsize Published in Commun Nonlinear Sci Numer Simulat 103 (2021) 106003. $\,\,$ \tcm{DOI:10.1016/j.cnsns.2021.106003}}\\
\tcb{{\scriptsize arXiv:2012.14333v3}}}
\end{abstract}
\maketitle

\noindent {\bf I. INTRODUCTION}

\medskip

It is well established that the size evolution of unstable, spherical cavitation bubbles in 3-dimensions with  surface tension is  governed by the well-known RP equation \cite{Lord,Pless,Prosp}
\be
\rho_w\Big(RR_{TT}+\df{3}{2}{R_T}^2\Big)=\Delta P-\frac {2 \sigma}{R}~, \label{eq1}
\en
where $\rho_w$ is the density of the water, $R(T)$ is the radius of the bubble, $\Delta P= p-P_\infty$ is the pressure drop between the uniform pressure inside  the bubble and the
external pressure in the liquid at infinity (hydrostatic and sound field for example), and $\sigma$ is the surface tension of the bubble. For our analysis, we will assume an internal pressure  proportional to the external pressure, i.e., $p=(k+1)P_\infty$, which gives   $\Delta P= kP_{\infty}$ according to \cite{RBW}. 
In the simpler form with only the pressure difference in the right hand side, Eq.~(\ref{eq1}) was first derived by Rayleigh \cite{Lord}, but it was only  in 1949 that Plesset developed the form (\ref{eq1}) of the equation and applied it to the problem of traveling cavitation bubbles \cite{Pless}.

\ms

On the other hand, we can  extend the RP equation to study the evolution of the cell wall of  microorganisms such as bacteria and viruses,
by the inclusion of  an additional term  that accounts  for the  bending pressure of the thin outer shell.
However, the effects of mechanical properties of the outer shell in controlling and maintaining the
sizes of microorganisms  are not well known.
Because the  elastic energy per unit area of bending a thin shell is proportional to
the square of the curvature \cite{MKT}, the extended  RP equation (ERP) can be modified to include this  additional bending pressure term $p_b = Y h^2/R^2$ of the thin  outer shell of elastic modulus $Y$, and thickness $h$ coating the cell to read
\be
\rho_w\Big(RR_{TT}+\df{3}{2}{R_T}^2\Big)=\Delta P-\frac {2 \sigma}{R}+ \frac{Yh^2}{R^2}~. \label{eq2}
\en
Typical fixed values that we will use  are  $\rho_w=10 ^3$~ kg/m$^3$, $R_0=10^{-6}$~m, $P_\infty=101325$ ~Pa, $h=3 \times 10^{-9}$~m, while $k$ varies in the interval $[-1,0)$, noting that the case $k=-1$ corresponds to zero internal  pressure, while the case $k=0$ corresponds to a zero pressure drop between the interior and exterior of cell walls. The values for Young's modulus and surface tension $\sigma$ are allowed to vary in the ranges  $Y \in [2\times 10^8, 5\times 10^8]$~Pa and $\sigma \in [1\times 10^{-4}, 2 \times 10^{-2}]$~ N/m.
In choosing these values, we have been guided by data mentioned in the literature. In the decade-old short review
``Physical Virology" by Roos et al. \cite{RBW} 
it is mentioned that ``viral shells have effective Young's moduli ranging from that of polyethylene to that of plexiglas", i.e., from 200 to 300 MPa up to $\sim$ 3 GPa, although here we will not consider shells stiffer than 500 MPa in bending modulus. On the other hand, a statistical analysis for more than one hundred species of small viruses performed 
 by Lo\v sdorfer Bo\v zi\v c et al. \cite{LB2013} 
 provides a mean thickness of viral capsids of $\sim$ 3 nm.
 A surface tension on the surface of the capsids is mentioned in \cite{RBW}
 as generated by the osmotic pressure of the genome material inside the capsids. 

In this paper, we find analytical solutions of the  ``bubbles with shell"
model as expressed by Eq.~\eqref{eq2}. 

\bs

\noindent {\bf II. INTEGRATING FACTOR AND INTEGRATION VIA WEIERSTRASS EQUATION}

\medskip

To solve \eqref{eq2} we will use the initial conditions $R(0)=R_0$, and $R_T(0)=0$.
We further introduce nondimensional variables given by $R=R_0u$, and $T=T_c t$.
Consequently, (\ref{eq2}) becomes
\be
uu_{tt}+\df{3}{2}{u_t}^2=\frac{{T_c}^2}{{R_0}^2\rho_w}\left(kP_{\infty}-\frac {2 \sigma}{R_0u}+ \frac{Yh^2}{{R_0}^2u^2}\right)~, \label{eq3}
\en
subject to new initial conditions given by  $u(0)=1$ and $u_t(0)=0$.
The collapse (Rayleigh) time $T_c$, used for the non-dimensional analysis, of the vacuous bubble ($k=-1$)
of $1$~$\mu$m in radius, was given by Mancas and Rosu \cite{Man}
\be \label{eq4}
T_c=
\df{\Gamma(\frac{5}{6})}{\Gamma(\frac{4}{3})}\sqrt{\frac{\pi}{6}}R_0\sqrt{\df{\rho_w}{P_\infty}}=0.914681~R_0\sqrt{\frac{\rho_w}{P_\infty}}=0.914681 \times 9.934401\times 10^{-8}~ {\rm sec}.=9.08681 \times 10^{-8} ~{\rm sec}~.
\en
Furthermore, we let $\frac{{T_c}^2P_\infty}{{R_0}^2\rho_w}=\xi^2$, where $\xi=0.914681$ is an universal constant known as the three-dimensional Rayleigh factor \cite{Kud}, and using the notation $\frac{\xi^2}{P_\infty}\frac{2\sigma}{R_0}=\gamma$, and $\frac{\xi^2}{P_\infty}\frac{Yh^2}{{R_0}^2}=\alpha$ (also noticing that the quotients
$\gamma/\xi^2=p_{\sigma}/P_{\infty}$, and $\alpha/\xi^2=p_{\alpha}/P_{\infty}$ are pressure quotients of the initial size with surface pressure  $p_\sigma= 2\sigma/R_0$, and bending pressure $p_\alpha=Yh^2/{R_0}^2$), we write (\ref{eq3}) in the form
\be
uu_{tt}+\df{3}{2}{u_t}^2=k\xi^2-\frac{\gamma} {u}+\frac{\alpha}{u^2}~. \label{eq5}
\en
For the values of ranges of parameters described above, we obtain $\gamma \in [0.0016514,0.330281]$, $\alpha \in [0.0082893, 0.0371566]$, surface pressure $p_\sigma \in [2 \times 10^2, 4 \times 10^4]$~ Pa, and bending  pressure $p_\alpha \in [1.8 \times  10^3, 4.5 \times 10^3]$~ Pa.

By multiplying   \eqref{eq5}   by the  integrating factor $2u^2u_t$, we have the conservation form
\be
\dfrac{d}{dt} \left[u^3 {u_t}^2-\frac{2k\xi^2}{3}{u^3}+\gamma{u^2}-2\alpha u\right]=0~,\label{eq7}
\en
so that
 \be
{u_t}^2=\frac{2k \xi^2 }{3}-\frac{\gamma}{u}+\frac{2 \alpha }{u^2}+\frac{c_1}{u^3}\label{eq8},
\en
where $c_1$ is an integration constant that varies linearly with respect to the surface tension $\sigma$ and Young's modulus $Y$. Using the two initial conditions, this constant is
\be \label{eq9}
c_1(\alpha,\gamma)=-2 \alpha +\gamma -\frac{2 k\xi^2}{3}.
\en

For the empty cavity ($k=-1$) without surface tension or bending pressure, $\gamma=0$ and $\alpha=0$, we have $c_1=\frac{2 \xi^2}{3}$, which reduces \eqref{eq8} to
\be \label{eq10}
{u_t}^2=\frac{2\xi^2}{3}\left(\frac{1}{u^3}-1\right).
\en

The solution of this equation is found by inversion of  the integral
\be
t(u)=\frac {1}{\xi}\sqrt{\frac{3}{2}}\int_{u}^1\frac{w^{3/2}dw}{\sqrt{1-w^3}}~, \label{eq11}
\en
which   in parametric form becomes
\be
t(u)=\frac{2}{5\xi} \sqrt{\frac{3}{2}} \left[\frac{\sqrt{\pi } \Gamma \left(\frac{11}{6}\right)}{\Gamma \left(\frac{4}{3}\right)}-u^{5/2} \, _2F_1\left(\frac{1}{2},\frac{5}{6};\frac{11}{6};u^3\right)\right]~. \label{eq12}
\en

\begin{figure}[h!]
\centering
\includegraphics[scale=0.95]{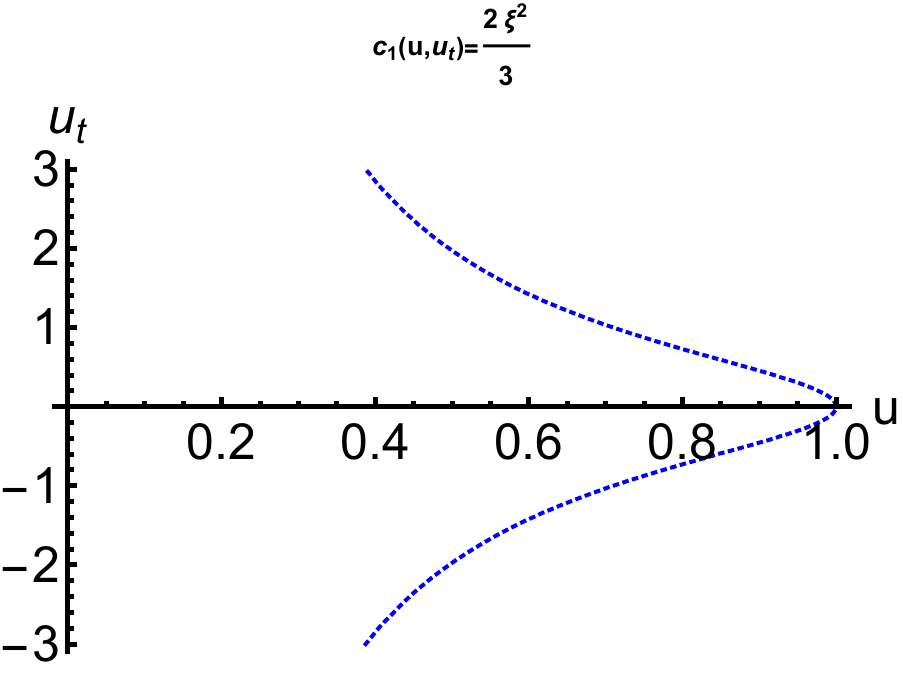}
\includegraphics[scale=0.95]{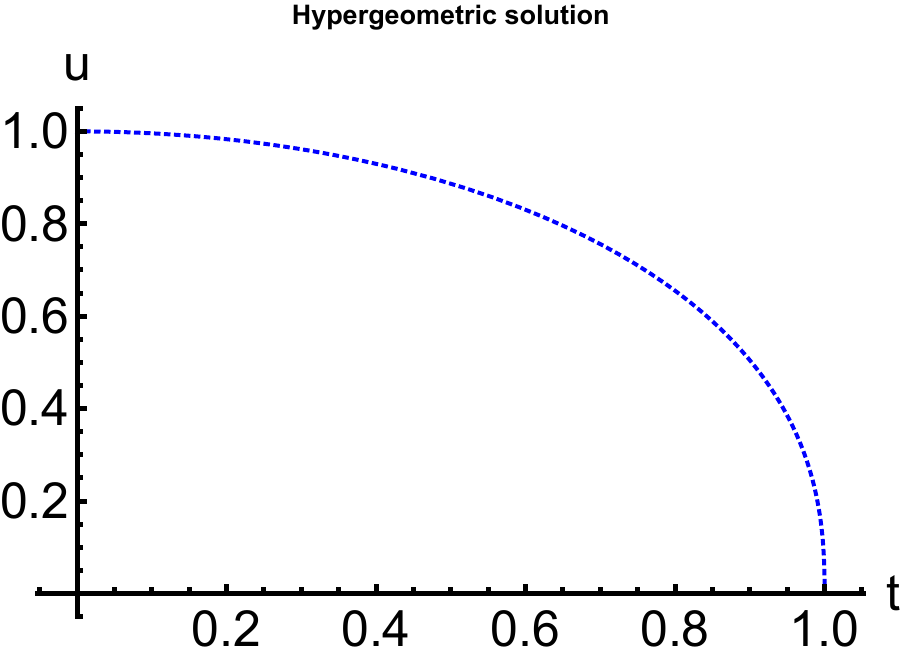}
\caption{\label{figure0} The phase portrait of (\ref{eq10}) and the corresponding parametric hypergeometric solution for the case $\gamma=0$ and $\alpha=0$,
{\em i.e.}, $(Y_0,\sigma_0)=(0,0)$, from  (\ref{eq12}).}
\end{figure}
Notice that the collapse time is obtained  from \eqref{eq11} by setting  $u=0$  which give the Rayleigh factor  $\xi=\df{\Gamma(\frac{5}{6})}{\Gamma(\frac{4}{3})}\sqrt{\frac{\pi}{6}}$.

The phase portrait of (\ref{eq10}) and the parametric hypergeometric solution for the case $\gamma=0$ and $\alpha=0$ from (\ref{eq12}) are displayed
in Fig.~\ref{figure0}.

\ms

 The relative contribution of surface tension and bending pressure determines the bending- and the tension-dominated regimes in which biological cells may be found. Arnoldi et al. \cite{Arnoldi2000} distinguished these regimes by the quotient of the corresponding free energies.
   However, the same quotient emerges as the  ratio  of the two constants $\alpha$ and $\gamma$ given by the parameter $q\equiv 2\alpha/\gamma=Yh^2/R_0\sigma>0$, which in this paper takes the values in the interval $[0.225,18]$.  When $q<1$ then  $\gamma>2 \alpha$ so we have the tension-dominated regime where $\sigma>Yh^2/R_0$. This is the regime of high surface tension which is characterized by irreversible deformation of the surface molecular layer. This region is analogous to the plastic deformation region. On the other hand, when  $q>1$ then  $\gamma<2 \alpha$ and we have the regime where $\sigma<Yh^2/R_0$ which is the region when microorganisms recover their shape after the external stress has been removed. This region is the elastic deformation region. At the boundary between the two regions, $q=1$, there is the critical radius $R_c=Yh^2/\sigma$ which plays an important role in determining the size of the microorganisms, which is the case that corresponds to the mean $\bar u$, special case II in equation (\ref{eq27a}) below.
In each of the plastic and elastic regions analytical solutions will be found as given in equation (\ref{eq27a}) by the maximum $u_M$ and minimum $u_m$, respectively. For the interested readers, we point out that in the case of filamentous (cylindrical) bacteria, the different regime parameter $\xi=pr/Yh$, but still in the form of a quotient of energies, has been introduced by Amir et al. \cite{Amir2014}.

\ms

The approach to integrate  $\eqref{eq8}$ is to transform it into an equation in which the right hand side is a cubic or quartic polynomial in $u$.
Namely, we will use the Sundman transformation
$dt=u^2d \tau$ where $\tau$ is the new independent variable which gives the Weierstrass elliptic equation
 \be
{u_\tau}^2=\frac{2k \xi^2}{3}u^4-\gamma u^3+2 \alpha u^2+c_1u\equiv Q(u)\label{eq19}.
\en

\ms

It is well known \cite{Wei, Whi, AS} that the solutions $u(\tau)$ of
\begin{equation} \label{eq20}
{u_\tau}^2=\mathrm{A}_4 u^4+4 \mathrm{A}_3 u^3+ 6 \mathrm{A}_2 u^2+ 4 \mathrm{A}_1 u+\mathrm{A}_0,
\end{equation}
can be expressed in terms of Weierstrass elliptic functions $\wp(\tau;g_2,g_3)$, which is a solution to
\begin{equation}\label{eq21}
{\wp_\tau}^2=4 \wp^3-g_2 \wp -g _3~,
\end{equation}
via the transformation
\begin{equation}\label{eq22}
u(\tau)=\hat u+\frac{ \sqrt{Q(\hat u)}\wp_\tau(\tau+\tau_0;g_2,g_3)+\frac 1 2 {Q}_u(\hat u)\Big[\wp(\tau+\tau_0;g_2,g_3)-\frac {1}{24}{Q}_{uu}(\hat u)\Big]+\frac{1}{24}Q(\hat u){Q}^{(3)}(\hat u)}{2\Big[\wp(\tau+\tau_0;g_2,g_3)-\frac{1}{24}{Q}_{uu}(\hat u)\Big]^2-\frac{1}{48}Q(\hat u)Q^{(4)}(\hat u)}~,
\end{equation}
where $\hat u$ can be taken not necessarily as a root of $Q(u)$, and $g_2,g_3$ are elliptic invariants of $\wp(\tau)$, given by 
\begin{equation}\label{eq23}
\begin{aligned}
	g_2&=\mathrm{A}_4 \mathrm{A}_0 -4 \mathrm{A}_3 \mathrm{A}_1 +3 {\mathrm{A}_2}^2=\frac{\alpha ^2}{3}+\frac{ c_1 \gamma }{4}~,\\
	g_3 &=\mathrm{A}_4 \mathrm{A}_2 \mathrm{A}_0 +2 \mathrm{A}_3 \mathrm{A}_2 \mathrm{A}_1 -\mathrm{A}_4 {\mathrm{A}_1}^2-{\mathrm{A}_2}^3-{\mathrm{A}_3}^2\mathrm{A}_0=-\frac{1}{216} \left(8 \alpha ^3+9k {c_1}^2 \xi^2 +9c_1 \alpha  \gamma \right)~.\\
\end{aligned}
\end{equation}
These invariants are components of the modular discriminant
\begin{equation}\label{eq24}
\Delta={g_2}^3-27 {g_3}^2=-\frac{{c_1}^2}{192}  \left(-3 \alpha ^2 \gamma ^2+9 {c_1}^2 k^2 {\xi} ^4-3c_1 \gamma ^3 +18 c_1 k \alpha   \gamma {\xi}^2 +16k \alpha ^3 {\xi}^2\right)
\end{equation}
and together are used to classify the solutions of \eqref{eq19}.
In particular, choosing $\hat u=0$, which is a root of $Q(u)=0$, the general solution \eqref{eq22}  takes the much simpler form
\begin{equation}  \label{eq25}
u(\tau)=\frac{Q_u(0)}{4 \wp (\tau+\tau_0; g_2,g_3)-\frac{Q_{uu}(0)}{6} }=\frac{A_1}{\wp (\tau+\tau_0; g_2,g_3)-\frac{A_2}{2}}~.
=\frac{c_1}{4\wp (\tau+\tau_0; g_2,g_3)-\frac{2 \alpha}{3}}~.
\end{equation}
This solution can also be explained by letting $u(\tau)=\frac{1}{v(\tau)}$ in \eqref{eq20} which gives the Weierstrass equation
\be
{v_\tau}^2=A_4 + 4 A_3 v+6 A_2v^2+4 A_1 v^3~,
\label{eq29r}
\en
which is
\be
{v_\tau}^2=\frac{2k\xi^2}{3} -\gamma  v+2 \alpha v^2+\left(-2 \alpha +\gamma -\frac{2 k \xi^2}{3}\right) v^3~.
\label{eq38}
\en
The standard form of (\ref{eq29r}) given by \eqref{eq21} can be found for $A_1 \ne 0 $  by  the linear transformation
\be  v(\tau)=\frac{1}{A_1}\left(\wp(\tau;g_2,g_3)-\frac{A_2}{2}\right)=\frac{1}{c_1}\left(4\wp (\tau+\tau_0; g_2,g_3)-\frac{2\alpha}{3}\right)
\label{eq30}
\en
yielding \eqref{eq25}.
Using  the initial conditions together with \eqref{eq9}, the constant $\tau_0$ can be found numerically by root finding methods from the equation
\be \label{eq26}
\wp (\tau_0; g_2,g_3)=\frac{3\gamma-2k\xi^2-4\alpha}{12}~,
\en
and thus the general solution to \eqref{eq19} in parametric form is
\begin{equation}\label{eq27}
\begin{aligned}
	u(\tau) &=\frac{3 \alpha +k \xi^2 -\frac {3\gamma}{2} }{ \alpha -6 \wp \left[\tau+\tau_0 ;\frac{\alpha ^2}{3}+\frac{\gamma }{4}  \left(-2 \alpha +\gamma -\frac{1}{3} 2 k \xi ^2\right) , -\frac{1}{216} \Big(4 \alpha -3 \gamma +2 k \xi ^2\right) \left(\alpha  (2 \alpha -3 \gamma )+2 k^2 \xi ^4+k \xi ^2 (8 \alpha -3 \gamma )\Big)\right]}~,  \\
	t(\tau) &=\int_0^\tau u^2(\zeta) d\zeta~.\\
\end{aligned}
\end{equation}

\noindent {\bf III. THE VACUOUS ($k=-1$) SHELL SOLUTIONS}

\ms

We set now $k=-1$ to present the ideal vacuous solutions of the Rayleigh-Plesset equation with a bending term.
Firstly, we select four sets of values of the parameters that we call
minimum, maximum, extreme zero values, and average values of Young's modulus and surface tension as presented in Table~I. 

\begin{center}
\begin{table}[h!]
{\footnotesize
 \noindent
\caption{{\small The numerical values of the parameters used in the phase portraits depicted in Fig.~\ref{figure1}.}}
\label{table1}
\medskip
\noindent\hfill
\begin{tabular}{|c||c||c||c||c|c|c|c|c|}
\hline
values for the parameters & $Y$ ~[Pa] & $\sigma$ ~[N/m] & $R_c$ ~[m]& $\gamma$ &  $\alpha$ & $c_1$ & $q$ & $u$ \\
\hline
\hline
minimum & $2 \times 10^8$ & $1.8 \times 10^{-5}$ & $10^{-5}$& $0.0016514$ & $0.0148626$ & $0.529687$ & $18$ & $u_m$ \\
\hline
maximum & $5 \times 10^8 $ & $2 \times 10^{-2}$ & $2.25 \times 10^{-7}$ & $0.330281$ & $0.0371566$ &  $0.831729$ & $0.225$ & $u_M$ \\
\hline
special I (extreme) & $0$ & $0$ & undefined & $0$ & $0$ & $0.557761$ & undefined & $u_0$ \\
\hline
special II (mean) &  $1.005 \times 10^{8}$ & $1.005 \times 10^{-2}$ &  $10^{-6}$& $0.165966$ & $0.0082893$ & $0.557761$ & 1 & $\bar u$ \\
\hline
\end{tabular}
\hfill}
\end{table}
\end{center}

For those values of the parameters, we obtain the following analytic solutions
\begin{equation}\label{eq27a}
\begin{aligned}
	u_m(\tau) &=\frac{0.132422}{\wp (\tau +3.30706;0.000292314,0.00977999)-0.0024771}~,\\
	u_M(\tau) &=\frac{0.203432}{\wp (\tau +2.72283;0.0676499,0.0226648)-0.00619276}~,\\
	u_0(\tau)&=\frac{\pi  \Gamma \left(\frac{5}{6}\right)^2}{36 \Gamma \left(\frac{4}{3}\right)^2 \wp \left(\tau +3.25193;0,\frac{\pi ^3 \Gamma \left(\frac{5}{6}\right)^6}{11664 \Gamma \left(\frac{4}{3}\right)^6}\right)}~,\\
	\bar{u}(\tau)&=\frac{0.13944}{\wp (\tau +3.15848;0.0254377,0.0105037)-0.0138305}~.
\end{aligned}
\end{equation}
Phase portraits of the elliptic Weierstrass equation \eqref{eq19}
for constant $c_1$ and the corresponding solutions of \eqref{eq27a} are displayed in Fig.~\ref{figure1}.

\begin{figure}[h!]
\includegraphics[scale=0.9]{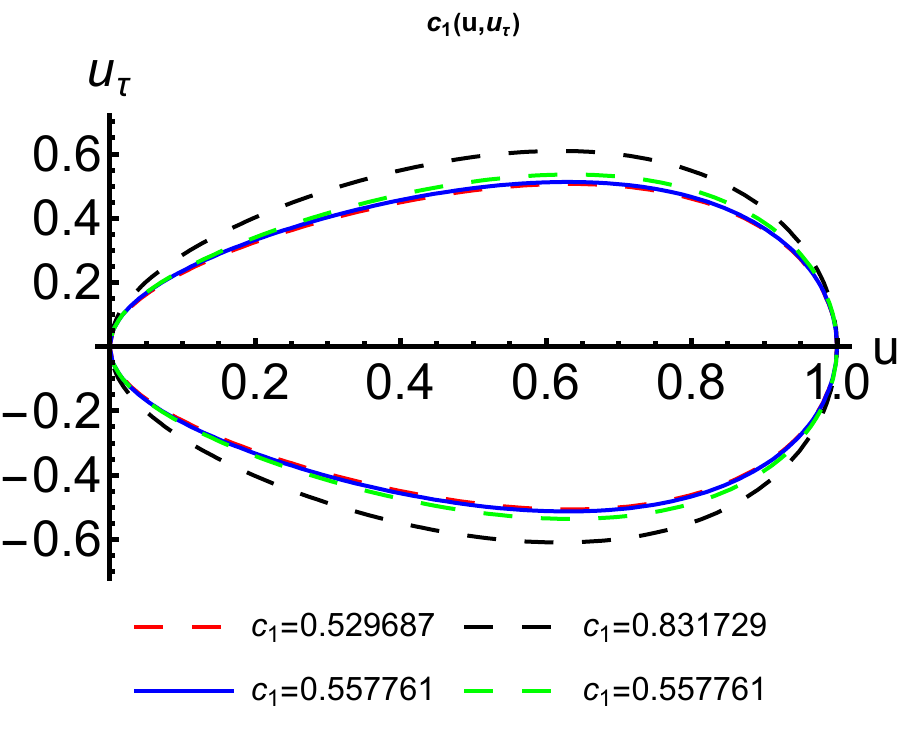}
\includegraphics[scale=0.9]{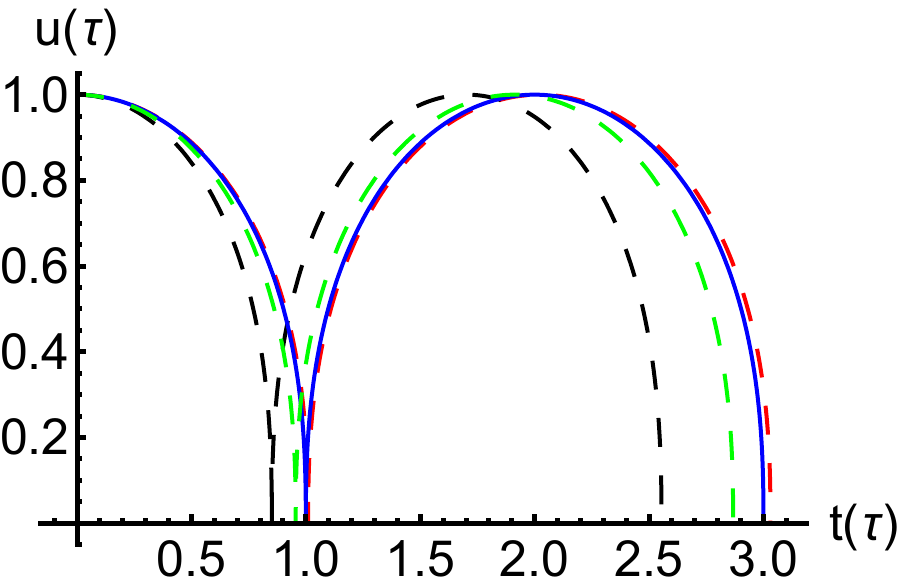}
\caption{\label{figure1} The phase portrait  of  (\ref{eq19}) which indicates periodic solutions given by \eqref{eq27a}. $c_1=0.529687$ and $c_1=0.831729$ correspond to minimum $(Y_m,\sigma_m)$ and maximum $(Y_M,\sigma_M)$ values, respectively. The two special cases of  $c_1=0.557761$ correspond to $(Y_0,\sigma_0)=(0,0)$, and  $(\bar Y,\bar \sigma)$. For $(Y_0,\sigma_0)=(0,0)$, the hypergeometric  solution can be parameterized  in terms of the $\wp$  elliptic function given by $u_0 (\tau)$.}
\end{figure}

\noindent {\bf A. Cnoidal solutions}

\medskip

This type of periodic solutions is obtained for the lemniscatic case $g_3=0$ which gives $\alpha=\frac{1}{4} (2 \xi^2 +3 \gamma )$, and is equivalent to $q=\frac 3 2 + \frac{\xi^2}{\gamma}$. In this case, $c_1=-\frac 1 6 (2 \xi^2+3 \gamma)$, $g_2=\frac{1}{48} (2 \xi^2 +\gamma ) (2 \xi^2 +3 \gamma )$, the roots of $Q(v)$ are real, and \eqref{eq29r} can be factored as
\be
{v_\tau}^2=-\frac{1}{6} (v-1) \left[-4 \xi^2 +v(2 \xi^2+3 \gamma)(v-2)\right]~.
\label{eq50}
\en
These real roots  are
$$
e_3=1-\frac{2 \xi^2 +\gamma }{\sqrt{\frac{4 \xi ^4}{3}+\frac{8 \xi^2  \gamma }{3}+\gamma ^2}}~,
\quad e_2=1~, \quad e_1=1+\frac{2 \xi^2 +\gamma }{\sqrt{\frac{4 \xi^4}{3}+\frac{8 \xi^2  \gamma }{3}+\gamma ^2}}~,
$$
and  although the Weierstrass unbounded function given by (\ref{eq21}) has poles aligned on the real axis of the $\tau-\tau_0$ complex plane, we can choose $\tau_0$ in such a way to shift these poles a half of period above the real axis, so that the $\wp$ elliptic function reduces to the Jacobi elliptic function given by
\begin{equation}\label{eq50}
\wp(\tau;g_2,0)=e_3+(e_2-e_3)\mathrm{sn}^2[\sqrt{e_1-e_3}(\tau+\tau_0);m]=-\frac{\sqrt{g_2}}{2}\mathrm{cn}^2\left[\sqrt[4]{g_2}(\tau+\tau_0);\frac{1}{\sqrt 2}\right]~,
\end{equation}  with elliptic modulus $m=\sqrt{\frac{e_2-e_3}{e_1-e_3}}.$ Thus, the solutions \eqref{eq30}
reduce to
\be
v(\tau)=1-\frac{24}{\xi^2+3\gamma}\,\wp (\tau; g_2,0)~.
\en
For the lemniscatic case, this solution is  obtained using the transformation (\ref{eq50}) to cnoidal waves, and it  becomes
\be
v(\tau)=1+\frac{2 \xi^2 +\gamma} 
{\sqrt{\frac{4 \xi^4}{3}+\frac{8 \xi^2 \gamma}{3}+\gamma ^2}}
\mathrm{cn}^2\left[\frac{\sqrt[4]{(2 \xi^2 +\gamma ) (2 \xi^2 +3 \gamma )} }{2 \sqrt[4]{3}}(\tau +\tau_0);\frac{1}{\sqrt{2}}\right]~.\label{eq49}
\en
To satisfy the initial condition, $\tau_0$ is found numerically from
\be
\mathrm{cn}\left[\frac{\sqrt[4]{(2 \xi^2 +\gamma ) (2 \xi^2 +3 \gamma )} }{2 \sqrt[4]{3}}\tau_0;\frac{1}{\sqrt{2}}\right]=0~.\label{eq51}
\en
Choosing the  mean value of $\sigma=1.005 \times 10^{-2}$ N/m,  one can obtain $\gamma=0.165966$,
$Y=7.30417 \times 10^ 9$ Pa, $q=6.54104$, $c_1=-0.361864$, $\alpha=0.542795$, and $R_c=6.54104\times 10^{-6}$ m.
The resulting analytic  solution  
is
\be \label{eq30x}
u(\tau)=\frac{1}{1+1.59416 ~\text{cn}^2\left[0.537061 (\tau +3.88405);\frac{1}{\sqrt{2}}\right]}~.
\en
The plot of this solution together with its phase portrait is presented in Fig.~\ref{figure3} showing that in this case the bubble does not collapse.
\begin{figure}[h!]
\centering
\includegraphics[scale=0.9]{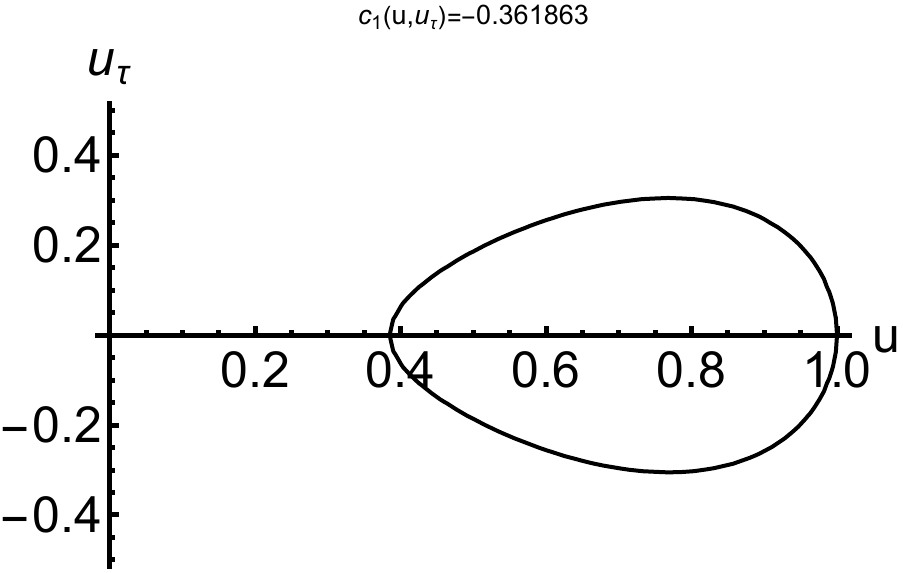}
\includegraphics[scale=0.9]{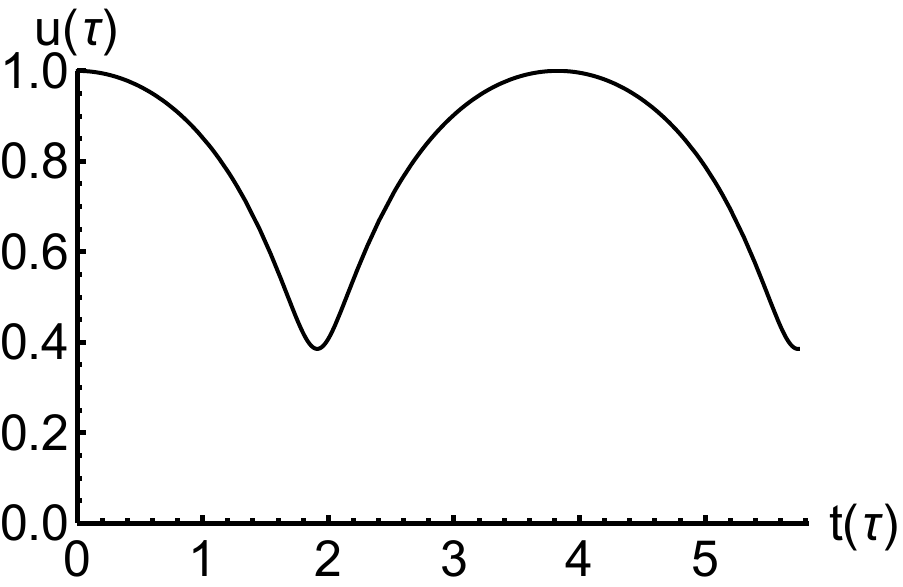}
\caption{\label{figure3}  The phase portrait from (\ref{eq19}) and the corresponding periodic parametric solution in terms of
Jacobi's elliptic function given by (\ref{eq30x}).
}
\end{figure}

\newpage


\noindent {\bf B. Degenerate cases}

\medskip

We now study the degenerate cases given by  $\Delta=0$ for which  \eqref{eq29r}  becomes
\be
{v_\tau}^2=\left(-2 \alpha+\gamma+\frac{2\xi^2}{3}\right)v^3+2 \alpha v^2-\gamma v-\frac{2\xi^2}{3}~. \label{eq31}
\en
In this case the discriminant factors as
\be
\Delta =\frac{(6 \alpha -2 \xi^2 -3 \gamma )^2 (\alpha -\xi^2 -\gamma )^2 \left(16 \alpha  \xi^2 -4 \xi^4-4 \xi^2  \gamma +3 \gamma ^2\right)}{1728}
\en
and  the solutions given by \eqref{eq30} simplify since  the Weierstrass $\wp$ function 
degenerates into trigonometric or hyperbolic elementary solutions.

\medskip

\noindent {\bf {i)} Trigonometric solutions}

\medskip

There are three possibilities for which $\Delta=0$.

\medskip

In the first case,  $\alpha=\xi^2+\gamma$ which is equivalent to  $q=2\left(1+\frac {\xi^2}{\gamma}\right)$
and  implies  $c_1=-\left(\frac{4 \xi^2 }{3}+\gamma\right)$. Then \eqref{eq31} has a double root at  $v=1$, which can be factored as
\be
{v_\tau}^2=-\frac{1}{3} (v-1)^2 [2 \xi^2 +(4 \xi^2 +3 \gamma ) v]~.
\label{eq32}
\en
The  solution is
\be
v(\tau)=1-\frac{3(2 \xi^2 +\gamma )}{4 \xi^2 +3 \gamma } \sec ^2\left[\frac{1}{2} \sqrt{2 \xi^2 +\gamma} \left(\tau+\tau_0\right)\right]~.
\label{eq33}
\en
However, this case  does not satisfy the initial condition $v(0)=1$, so it will be disregarded as nonphysical.

\medskip

Secondly, $\alpha=\frac{1}{16} \left(-\frac{3 \gamma ^2}{\xi^2 }+4 \xi^2 +4 \gamma \right)$  which is equivalent to $q=\frac 12+\frac{\xi^2}{2\gamma} -\frac{3\gamma}{8\xi^2}$, and
 gives $c_1=\frac{(2 \xi^2 +3 \gamma )^2}{24 \xi^2 }$. Then \eqref{eq31} has a simple root for $v=1$, which can be factored as
\be
{v_\tau}^2=\frac{1}{24 \xi^2 }(v-1) [4 \xi^2 +(2 \xi^2 +3 \gamma ) v]^2
\label{eq34}
\en
with solution
\be
v(\tau)=-\frac{4 \xi^2}{2 \xi^2 +3 \gamma }+\frac{3 (2 \xi^2 +\gamma )}
{2 \xi^2 +3 \gamma }\sec ^2\left(\frac{ \sqrt{2 \xi^2 +\gamma } \sqrt{2 \xi^2 +3 \gamma }}{4 \sqrt{2}\xi}\tau\right)~,
\label{eq35}
\en
which satisfies the initial condition $v(0)=1$.
The general solution to \eqref{eq19} in parametric form is
\begin{equation}\label{eq36}
\begin{aligned}
u(\tau)&=\frac{1}{1+A\tan^2(\theta  \tau)}\\  
 t(\tau) &= -\frac{1}{2\theta(A-1)}\Bigg[\frac{A \tan (\theta  \tau)}{1+A\tan^2(\theta  \tau)}+\frac{2\theta}{A-1}\tau+
\frac{(A-3) \sqrt{A}}{A-1} \tan ^{-1}\left(\sqrt{A} \tan (\theta  \tau)\right)
\Bigg]~,
\\
\end{aligned}
\end{equation}
where $A=1+\frac{4\xi^2}{2\xi^2+3\gamma}$, and $\theta=\frac{ \sqrt{2 \xi^2 +\gamma } \sqrt{2 \xi^2 +3 \gamma }}{4 \sqrt{2} \xi}$.
Choosing the  mean value of $\sigma=1.005 \times 10^{-2}$ N/m,  one can obtain
$\gamma=0.165966$, $Y=3.28985 \times 10^ 9$ Pa, $q=2.94613$, $c_1=0.234769$, $\alpha=0.244479$, and $R_c=2.94613\times 10^{-6}$ m.

Using these values, one finds $A=2.54136$ and $\theta=0.38621$.
The corresponding periodic trigonometric solution \eqref{eq36} and its phase portrait are presented in Fig.~\ref{figure4}.

\medskip

\noindent {\bf {ii)} Hyperbolic  solutions} \\
This case is found when $\alpha=\frac{1}{6} (2 \xi^2 +3 \gamma )$, which is equivalent to $q=1+\frac{2\xi^2}{3\gamma}$,
and  gives $c_1=0$. Then \eqref{eq31} is  factored as
\be
{v_\tau}^2=\frac{1}{3} (v-1) \left[2 \xi^2 +(2 \xi^2 +3 \gamma ) v\right]~,
\label{eq32}
\en
with solution
\be
v(\tau)=\frac{3\gamma}{2(2 \xi^2 +3 \gamma )} +\frac{4 \xi^2 +3 \gamma}{2(2 \xi^2 +3 \gamma )}
\cosh \left(\sqrt{\frac{2 \xi^2+3\gamma}{3}}\,\tau\right)~,
\label{eq33}
\en
which satisfies the initial condition $v(0)=1$.
The general solution to \eqref{eq19} in parametric form is
\begin{equation}\label{eq38}
\begin{aligned}
u(\tau)&=\frac{1}{1+2B\sinh^2 \left(\frac{\tilde{\theta}  \tau }{2}\right)}~,\\
	t(\tau) &= \frac{1}{\tilde{\theta}(2B-1)}\Bigg[\frac{B \sinh (\tilde{\theta}  \tau )}{1+2B\sinh^2 \left(\frac{\tilde{\theta} \tau}{2}\right)}+\frac{2B-2}{\sqrt{2 B-1}} \tan ^{-1}\left(\sqrt{2 B-1} \tanh \left(\frac{\tilde{\theta} \tau }{2}\right)\right)
\Bigg]~,\\
\end{aligned}
\end{equation}
where $B=\frac{1}{2}+\frac{\xi^2}{2\xi^2+3\gamma}$, and $\tilde{\theta}=\sqrt{\frac{2 \xi^2+3\gamma}{3}}$.
Choosing the  mean value of $\sigma=1.005 \times 10^{-2}$ N/m,  one can obtain
$\gamma=0.165966$, $Y=4.86994 \times 10^ 9$ Pa, $q=4.3607$, $c_1=0$, $\alpha=0.361864$, and $R_c=4.3607\times 10^{-6}$ m.

Using these values, then $B=0.885339$, and $\tilde{\theta}=0.850722$.
For these values, the plot of the hyperbolic solution \eqref{eq38} and its phase portrait are presented in Fig.~\ref{figure4}.

\begin{figure}[h!]
\centering
\includegraphics[scale=0.9]{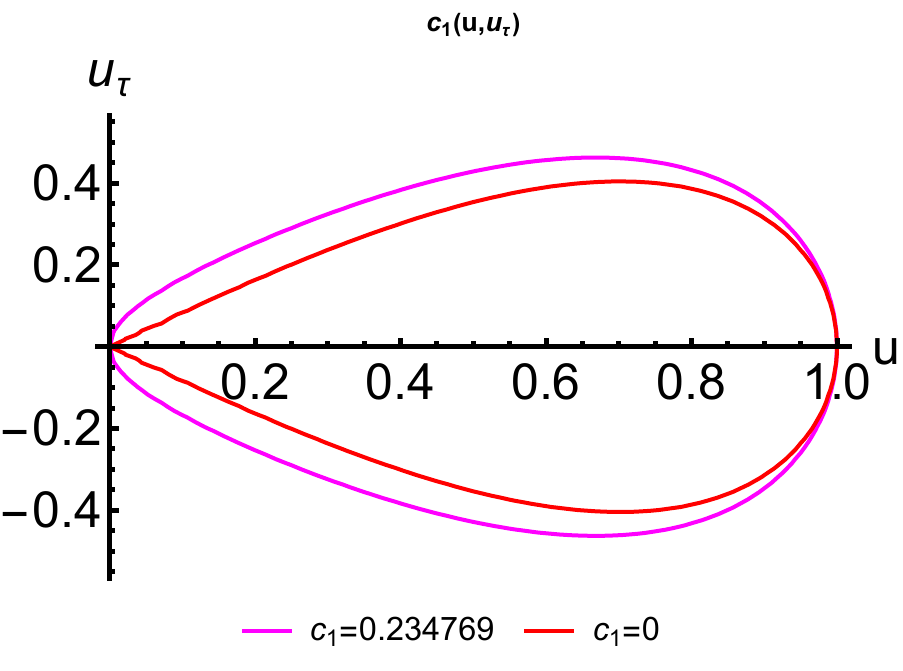}
\includegraphics[scale=0.9]{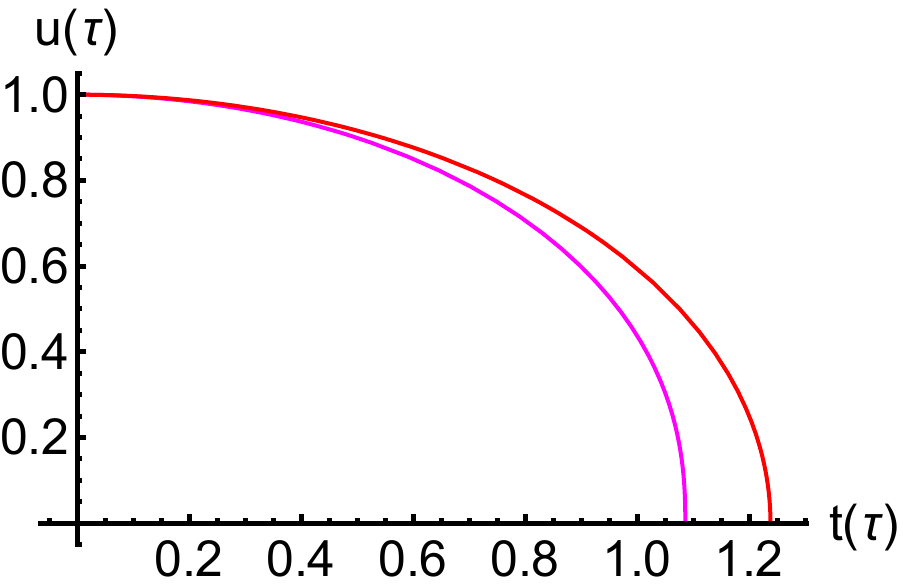}
\caption{\label{figure4}  The phase portraits from (\ref{eq19}) and the corresponding periodic trigonometric solution (\ref{eq36}) and hyperbolic solution (\ref{eq38}).}
\end{figure}

\noindent {\bf IV. SOLUTIONS OF THE RAYLEIGH-PLESSET WITH SHELLS EQUATION FOR $k\neq -1$}

\ms

For $k\neq -1$, the solutions can be considered more realistic because this case implies a non zero internal pressure.
In the range $k\in (-1,0)$, the solutions are still bound, and unless for a shift with respect to the origin they are not
really different from the vacuous solutions as shown by the plots presented in Fig.~\ref{figure5} for the case $k=-0.1$ of rational Weierstrass
solutions.

On the other hand, for strictly positive values of $k$, there are only unbounded solutions since the internal pressure is bigger than the outside one.
Plots of the $k=0.1$ case are presented in Fig.~\ref{figure6}. We surmise that these unbounded solutions correspond to the fundamental phenomenon of pressure-driven DNA ejection associated to the majority of the bacterial viruses and to some of the eukaryotic viruses \cite{Evilev13,Hanh13}.

We further notice that periodic solutions of cnoidal type are possible for the special value of $k$ given by
\begin{equation}
k\equiv k_{\rm cn}=\frac{\alpha\left[8\alpha^2+9c_1\gamma\right]}{(3c_1\xi)^2}
\end{equation}
and directly degenerate trigonometric solutions are possible if
\begin{equation}
k\equiv k_{\rm d}=\frac{-\alpha(8\alpha^2+9c_1\gamma)\pm \sqrt{\alpha^2(8\alpha^2+9c_1\gamma)^2-27{c_1}^2\gamma^2(\alpha^2+c_1\gamma)}}{(3c_1\xi)^2}~
\end{equation}
when the combination of the parameters is such that $k_{\rm d}\in (-1,0)$.
\begin{figure}[h!]
\includegraphics[scale=0.9]{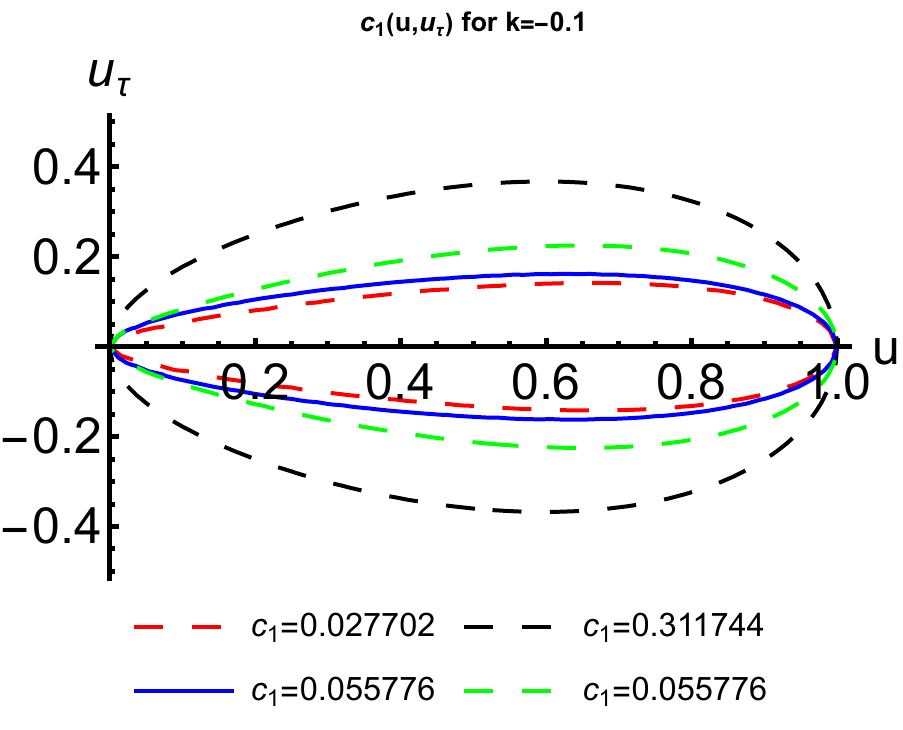}
\includegraphics[scale=0.9]{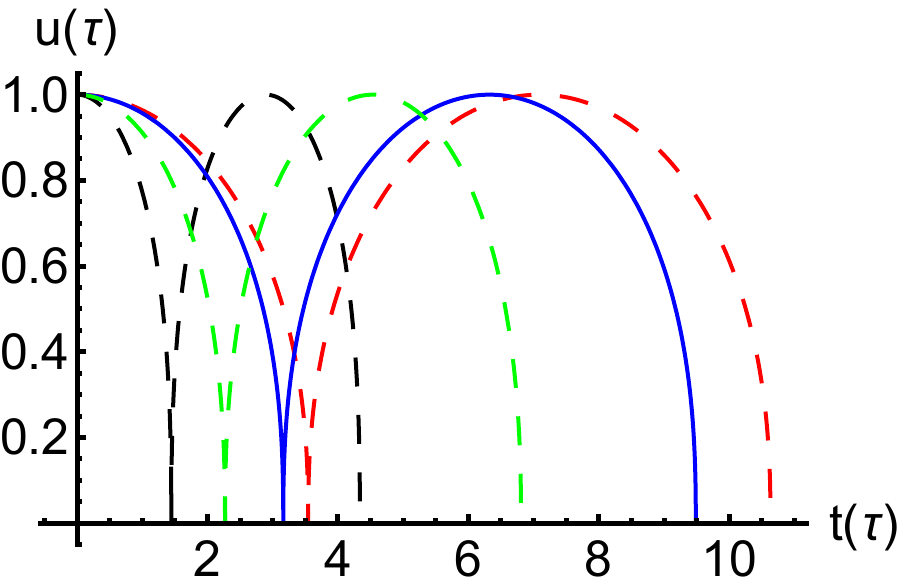}
\caption{\label{figure5}
The phase portrait and the corresponding periodic rational Weierstrass solutions 
 in the case $k=-0.1$. $c_1=0.027702$ and $c_1=0.311744$ correspond to minimum and maximum values, $(Y_m,\sigma_m)$ and $(Y_M,\sigma_M)$, respectively. The two special cases of  $c_1=0.055776$ correspond to $(Y_0,\sigma_0)=(0,0)$ and $(\bar Y,\bar \sigma)$, hypergeometric and mean cases, respectively.}
\end{figure}

\begin{figure}[h!]
\includegraphics[scale=0.9]{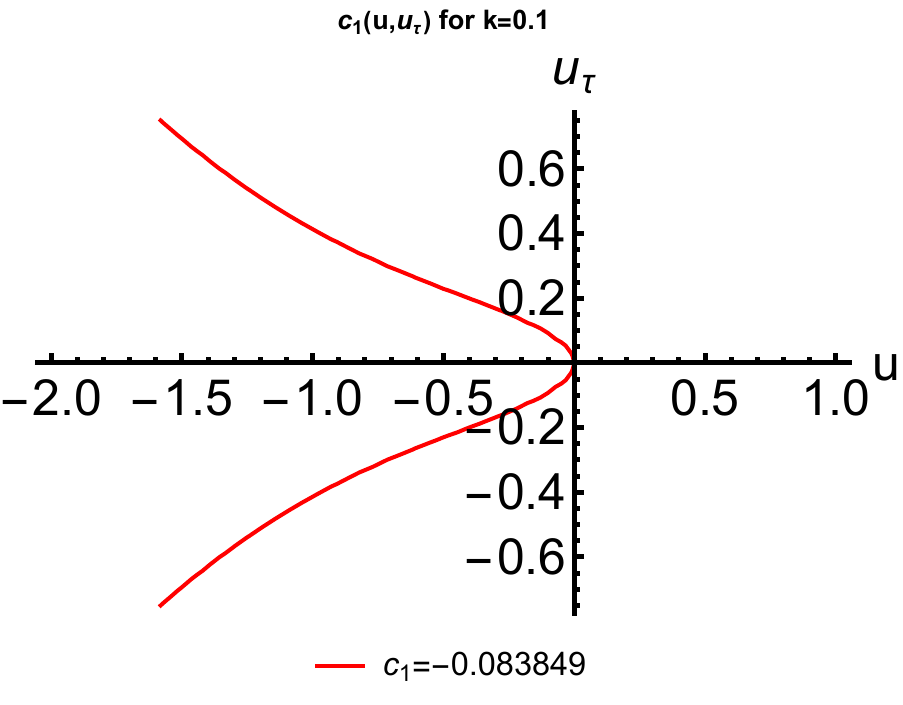}
\includegraphics[scale=0.9]{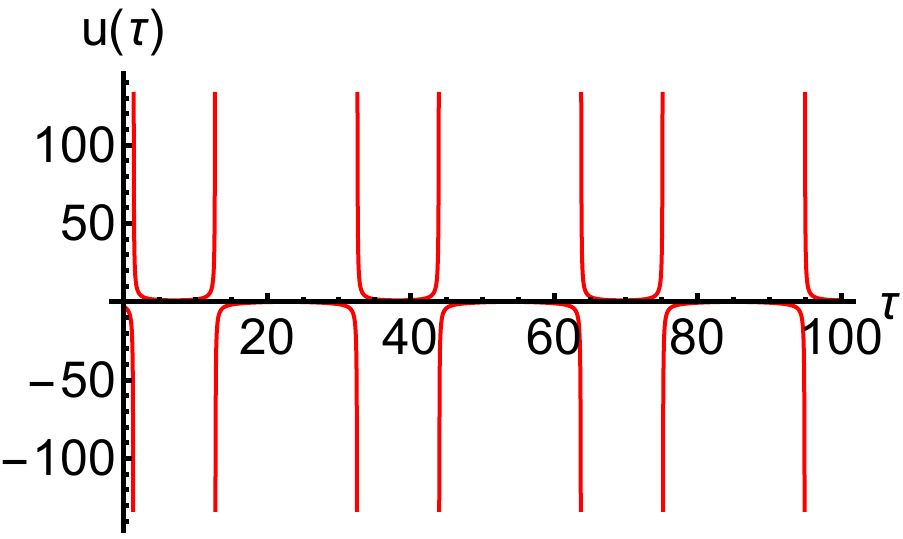}
\caption{\label{figure6} The unbound case for $k=0.1$ for which $c_1=-0.083849$.}
\end{figure}

\bs

\noindent {\bf V. CONCLUSION}

\medskip

In this paper, parametric solutions of the Rayleigh-Plesset equation extended with a term that takes into account the bending pressure due to the elasticity of a shell or capsule surrounding a liquid- or vapor-like substance have been obtained. The general method of Weierstrass elliptic equation using as evolution parameter the Sundman time has been employed. Particular cases that can be important in applications, such as cnoidal and modular-degenerate solutions, are also presented. The simpler, but more particular method using the Abel equation has been briefly described in the appendix. The quotients of the surface and bending pressures and the pressure of the background medium together with the Rayleigh collapse time are the other parameters that characterize the solutions displayed in this work.

\newpage

\noindent {\bf Declaration of competing interest}\\

The authors declare that they have no known competing financial interests or personal relationships
that could have appeared to influence the work reported in this paper.

\bigskip
\bigskip

\noindent {\bf Credit authorship contribution statement}\\

\noindent {\bf S.C.~Mancas}: Writing of initial version, Methodology, Calculations.

\noindent {\bf H.C.~Rosu}: Supervision, Validation, Calculations.

\noindent {\bf C.-C.~Hsieh}: Supervision, Project administration.

\bigskip

\noindent {\bf Acknowledgements}\\

We wish to thank the anonymous referees for their remarks that helped us to improve significantly this paper.


%
%



\bs

 \renewcommand{\theequation}{A\arabic{equation}}
  \setcounter{equation}{0}  

\noindent {\bf APPENDIX A: INTEGRATION  VIA ABEL\rq{}S EQUATION}

   \medskip

Proceeding as in Mancas and Rosu \cite{Man3}, the solutions to a general second order ODE of type
\begin{equation}\label{eq13}
u_{tt}+f_2(u) u_t +f_3(u)+f_1(u) { u_t}^2+f_0(u) {u_t}^3=0
\end{equation}
 can  be obtained via the solutions to Abel's equation  of the first kind (and vice-versa)
\begin{equation}\label{eq14}
\frac{dy}{du}=f_0(u)+f_1(u)y+f_2(u)y^2+f_3(u)y^3
\end{equation}
using the substitution
\begin{equation}\label{eq15}
u_t=\eta(u(t))~,
\end{equation}
which turns (\ref{eq13}) into the Abel equation of the second kind in canonical form
\begin{equation}\label{eq16}
\eta \eta_u+f_3(u)+f_2(u)\eta+f_1(u)\eta^2+f_0(u)\eta^3=0~.
\end{equation}
Using  the inverse transformation $\eta(u(t))= 1/y(u(t))$ 
of the dependent variable, (\ref{eq16}) becomes (\ref{eq14}) and viceversa.

\ms

In our case, by comparing \eqref{eq13} with \eqref{eq5}, we identify the nonlinear coefficients to be  $f_0(u)=0$, $f_1(u)=3/(2u)$, $f_2(u)=0$, and $f_3(u)= -k\xi^2/u+\gamma/u^2-\alpha/u^3$. Therefore Abel\rq{}s equation \eqref{eq14} simplifies to the Bernoulli equation
\be
\frac{dy}{du}=f_1(u)y+f_3(u)y^3~.\label{eq17}
\en
By one quadrature, this equation has the solution
 \be
y(u)=\pm\frac{u^{3/2}}{\sqrt{c_1+2 \alpha u-\gamma u^2+\frac{2 k\xi^2}{3}  u^3}}~,\label{eq18}
\en
and using the inverse transformation $1/y(u(t))=u_t$ together with  \eqref{eq15}, one can obtain \eqref{eq8}.

\end{document}